\documentclass[journal=apchd5,manuscript=article]{achemso}
\usepackage{amssymb} 

\usepackage[version=3]{mhchem} 
\usepackage{epstopdf} 

\usepackage{tabularx}
\usepackage{booktabs}
\newcommand{\tableheadline}[1]{\multicolumn{1}{c}{\small}}
\usepackage{caption}
\usepackage{subfig}

\usepackage{graphicx}
\usepackage{dcolumn}
\usepackage{bm}

\usepackage[utf8]{inputenc}
\usepackage[T1]{fontenc}
\usepackage{mathptmx}
\usepackage{wrapfig}
\usepackage{color}

\DeclareGraphicsExtensions{.pdf,.eps,.eps,.jpg,.mps}
\DeclareUnicodeCharacter{0305}{\overline}

{\title{Doping-Dependent Optical Response of a Hybrid Transparent Conductive Oxide/Plasmonic Medium\\}}

\author{Maria Sygletou}
\email{ sygletou@fisica.unige.it}
\affiliation{OptMatLab, Dipartimento di Fisica, Universit\`{a} di Genova, via Dodecaneso 33, I-16146 Genova, Italy}
\author{Stefania Benedetti}
\affiliation{CNR-Istituto Nanoscienze, via Campi 213/a, 41125 Modena, Italy}
\author{Alessandro di Bona}
\affiliation{CNR-Istituto Nanoscienze, via Campi 213/a, 41125 Modena, Italy}
\author{ Maurizio Canepa}
\affiliation{OptMatLab, Dipartimento di Fisica, Universit\`{a} di Genova, via Dodecaneso 33, I-16146 Genova, Italy}
\author{Francesco Bisio}
\affiliation{CNR-SPIN, C.so Perrone 24, I-16152 Genova, Italy}

\begin{document}
{\color{red}{This article may be downloaded for personal use only. Any other use requires prior permission of the author and ACS Publishing.  This article appeared in (M. Sygletou,The Journal of Physical Chemistry C (2022)) and may be found at (https://10.1021/acs.jpcc.1c07567).}}
\begin{abstract}

Understanding the interaction between plasmonic nanoparticles and transparent conductive oxides is instrumental to the development of next-generation photovoltaic, opto-electronic and energy-efficient solid-state lighting devices.
We investigated the optical response of hybrid media composed of gold nanoparticles deposited on aluminum-doped zinc oxide thin films with varying doping concentration by spectroscopic ellipsometry. 
The dielectric functions of bare AZO were addressed first, revealing doping-induced effects such as the band-gap shift and the appearance of free carriers.
In the hybrid media, a blueshift of the localized surface plasmon resonance of Au NPs as a function of increasing Al-doping of the substrate was observed, ascribed to the occurrence of a charge transfer between the two materials and the doping-dependent variation of the polarizability of the substrate.

\end{abstract}


\maketitle

\section{\label{sec:level1}Introduction}

Transparent conductive oxides (TCOs) have been extensively used in optoelectronics due to their visible-light optical transparency combined with low electrical resistivity,\cite{Ellmer,Naghdi2018ARO,KangReview2019} and for the possibility to realize so-called epsilon-near-zero (ENZ) materials.\cite{Gurung2020,Vezzoli2018,Naik8834}
Aluminum-doped ZnO (AZO) is one of such materials, based on an n-type semiconductor with direct wide-energy band gap, and cheaper than the commonly-used indium-tin oxide.
Aluminum doping enhances the conductivity of ZnO, making AZO suitable for applications in optoelectronic devices, as a transparent conductive component and ENZ material.\cite{ReviewAZO,Viktoriia2015}
AZO is cheap and easy to fabricate with various techniques, such as dc magnetron sputtering \cite{propertiessputtering}, atomic layer deposition\cite{ALD_AZO}, pulsed-laser deposition\cite{PLD_AZO,Gondoni_2012} etc. 
The optical properties of AZO can be tailored, to some extent, acting on the deposition parameters, like doping level \cite{Huali_2008,AZOdoping_ALD}, thickness\cite{Pradhan2014,ALD_AZO_thickness} or the substrate temperature.\cite{propertiessputtering,Qing_Geng_2008}\par
The combination of TCOs with plasmonic nanoparticles (NPs) allows the realization of new kind of hybrid systems \cite{Abb2011,AuNPs_AZO2011,Chew2015,Kravets2018,KangReview2019,Ag_Ceriumoxide2020,metalproperties2003,Malinsky2001} 
and metasurfaces\cite{Bruno2020}.
Among the various phenomena associated with the localized surface plasmon resonance (LSPR), there is growing interest in the excitation of so-called hot electrons, {\em i.e.} energetic carriers that play a role in a variety of plasmon-induced phenomena, such as photocatalysis, photosynthesis and more.\cite{Clavero2014b,kale2014}
When NPs are in contact with TCOs, hot electrons can be injected from the NPs into the TCO, by quantum tunneling through the Schottky barrier.\cite{Clavero2014b,Kodiyath2014,Catone2019} 
The injection of hot electrons causes a redistribution of charge carriers due to the Seebeck effect.\cite{Schroeders1991,Abb2011,Clavero2014b,Kodiyath2014,Viktoriia2015} 
Within a NP-TCO hybrid system, not only the plasmonic NPs can modify the electro-optical properties of the TCO but also the TCO can tune the plasmon resonance of the NPs, for example by modifying the TCO dielectric function {\em via} doping.
This type of hybrid plasmonic-TCO systems in which the optical and electrical response can be controlled, are extremely promising for the realization of innovative plasmo- and opto-elctronics devices for ultrafast and steady-state applications.\cite{Abb2011,plasmonicsolarcells2008,Zhang2010,Yang2010,Abb_2012,stratakis2013,Kumar2014,JVS_Sygletou2019, Clavero2014b,Kodiyath2014,HEreview,Kumar2016,Saavedra2016,Brongersma2015,Baffou2011,liu2018,smith2018,Ferrera2020,Bruno2020,Jun2013}

We based our investigations of these hybrid systems upon a very precise assessment of their dielectric characteristics (stand-alone TCO, and hybrid NP/TCO), providing solid foundations for assessing the overall optical response.  
We addressed the optical properties of Au NPs deposited on bare and Al-doped ZnO films by means of
spectroscopic ellipsometry (SE), taking the optical response of the bare TCO films as a starting point for the discussion. 
We covered a photon energy range broad enough to encompass both the ultraviolet interband transitions and the free-carrier range in the IR (including the ENZ region), thereby evaluating the full materials response.\par
\begin{figure}
		\includegraphics[width=9cm]{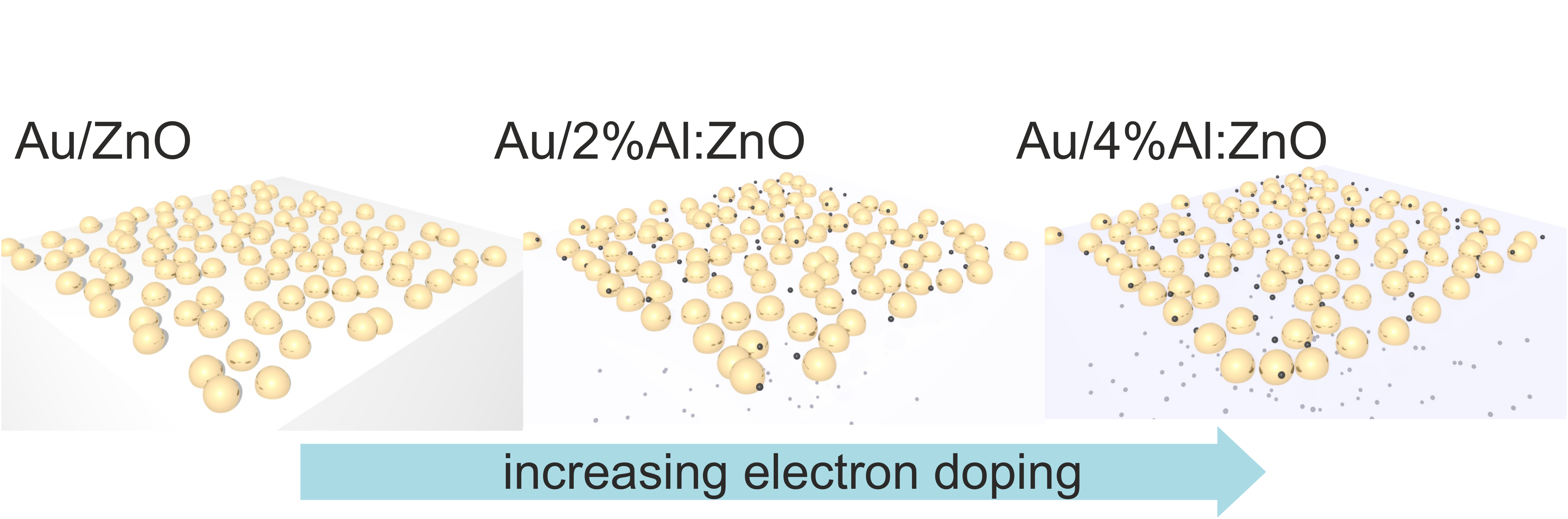}
		\caption{\label{scheme}Schematic of the evolution of the optical properties of NPs on top of TCO films as a function of the TCO's doping level.}
	\end{figure}

Increasing the Al doping in ZnO film from 0\% to 4\%, we observe mainly two classes of effects, respectively related to the stand-alone TCO and to the hybrid system.
Among the former, we mention the blueshift of the ZnO optical band gap \cite{Moss_1954} and the appearance of a plasma frequency in the near-IR due to the free carriers, whereas for the latter we observed a steady blueshift of the plasmon resonance of Au NPs for increasing doping concentration (see Figure \ref{scheme}). 
Whereas to the first order this blueshift can be  rationalized in terms of a variation of the effective dielectric environment of the NPs, the experimental evidence suggests that charge-transfer effects between the two materials have to come into play. 
In this work, we provide indications that there is a  charge transfer from the TCO to the metallic nanoparticles as a function of the doping concentration of the TCO, by combining SE and atomic force microscopy (AFM) techniques.

\section{\label{sec:level1}Experimental}

\subsection{\label{sec:level2}Sample preparation}
Previous studies report 4 at.\% AZO as the optimum Al concentration for obtaining the highest electrical conductivity of AZO.\cite{Benedetti2015,Benedetti2017}
Above this doping level, defects have been observed in the oxide related to the depopulation of its conduction band that reduce the number of free carriers and their mobility. Based on this, AZO films with nominal content of Al=2, 4 at.\% and bare ZnO films were deposited by magnetron sputtering on MgO (001) substrates. 
Before loading the substrates into the chamber, the substrates were stirred in acetone for 5 minutes, then ultrasonically cleaned in acetone and isopropanol in sequence for 5 minutes, and finally dried with nitrogen (N2, 99.999 \% purity). 
AZO films were obtained by co-deposition from a 3” RF magnetron source (ZnO) and a 3” DC magnetron source (Al) operating in confocal geometry approximately 15 cm far from the substrate. 
A constant 0.7 {\AA}/s ZnO deposition rate was reached at a RF power of 120 W. The DC power was varied in order to obtain the desired doping level, in the 0–4 at.\% range, defined as Al/(Al + Zn). During deposition the substrate temperature was set at 300$^{\circ}$C with a base pressure of ${1\cdot10^{-6}}$ mbar, in a $5\cdot10^{-3}$ mbar Ar gas atmosphere. A rotating sample holder was used to obtain uniform deposition. 
The doping level has been checked by energy-dispersive X-ray spectroscopy and hard X-ray photoemission spectroscopy (HAXPES) \cite{Benedetti2015}. The control of the dopant concentration in the sputtering method with RF on ZnO target and DC on Al target was rather difficult below 1\%. Furthermore the post-growth determination of the dopant concentration by quantitative analysis (like EDX) is limited at these low values, because of the reduced amount of Al and of the vicinity of Zn and Al signals, that can hardly be deconvolved when the amount of Al is too low. For these reasons we decided to investigate and compare films with a clear dopant concentration that could be determined by EDX with a good precision.
The morphopolgy of the surface of the AZO films was examined by AFM. The AFM images of bare ZnO and AZO films of different doping levels (2\% and 4\%), obtained in the same deposition conditions, are shown in Figure S1 of the Supplementary information (SI). 
The deposition parameters were set after a parametric study for the optimization of the surface roughness of bare ZnO films on MgO substrates. Indicative microscopy images of this study are presented in Figure S2 of the SI. The root mean square surface roughness of the bare ZnO film was around 1.5 nm, and it increased up to 2.2 nm for 2 at.\% AZO and up to 2.7 nm for 4 at.\% AZO film. 

\begin{figure}
		\includegraphics[width=9cm]{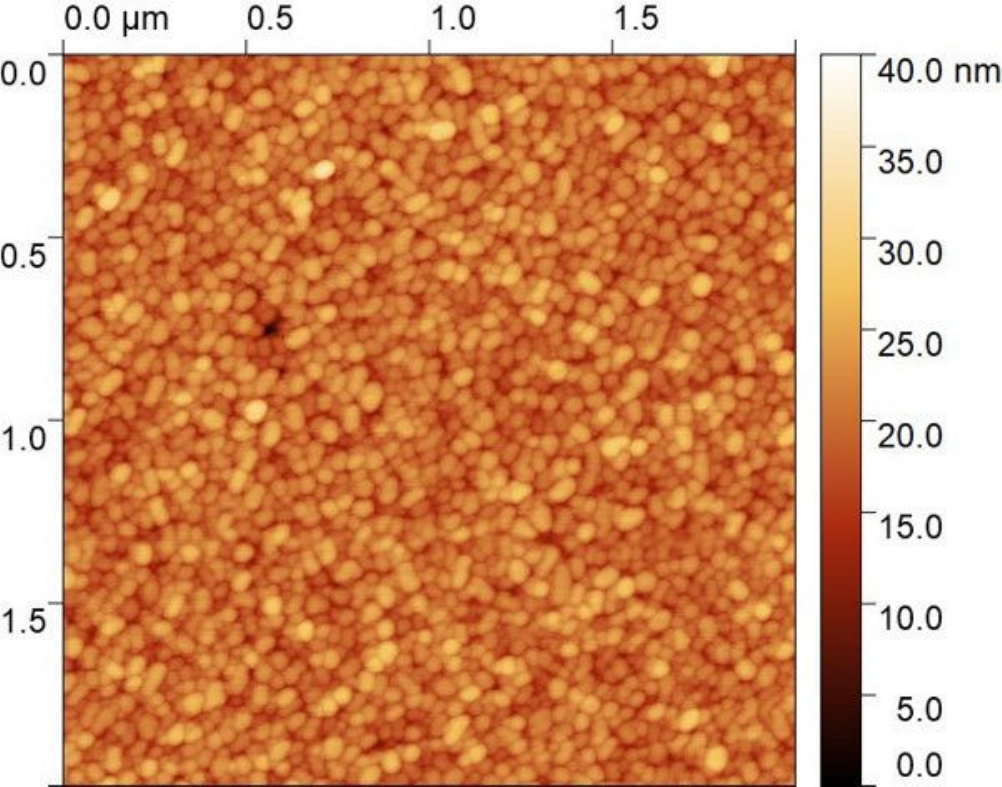}
		\caption{\label{AFM_2percento}AFM image of Au NPs deposited on top of 2 at.\% AZO film, grown on MgO substrate.}
	\end{figure}

Au NPs were deposited on AZO by molecular-beam epitaxy: 3 nm of Au deposited at normal incidence angle by molecular beam epitaxy at p$\approx10^{-9}$ mbar on the AZO/MgO substrate in a dedicated chamber. 
The Au/AZO/MgO system was annealed at $400^{\circ}$C in the deposition chamber, resulting in the formation of isolated NPs, according to AFM. The AFM image in Figure \ref{AFM_2percento} shows the surface morphology of Au NPs deposited on top of 2 at.\% AZO film. No significant differences are observed on the Au NPs grown on the different substrates (see Figure S1-bottom). The mean size of the NPs was around 20-25 nm.

\subsection{\label{sec:level2}Spectroscopic Ellipsometry}

SE is a very sensitive and non-destructive technique for investigating the optical response of materials, successfully applied to various TCOs and other systems of complex nanoscale morphology.
\cite{Al2O3_ZnO_2011,Zno_thickness_2014,Zollner2017,AZOdoping_ALD,Pradhan2014,propertiessputtering,Qing_Geng_2008,Huali_2008} 
It was perfomed by means of J.A. Woollam V-VASE ellipsometer (0.49-5.05 eV range, incidence angles of 60$^{\circ}$ and  65$^{\circ}$), under ambient conditions. 
SE is based on the measurement of the variation of the polarisation state of light reflected at non-normal incidence off the sample surface;
it yield the so-called ellipsometry angles $\Psi(\lambda$) and  $\Delta(\lambda$), defined by the equation $r_{p}/r_{s}=\tan \Psi\cdot e^{i\Delta}$, where $r_{p(s)}$  are the Fresnel reflection coefficients for $p(s)$-polarized radiation. \cite{Pal2018,Bundesmann2004,Toccafondi2014} 

From the optical point of view, the system was modelled as a stack of dielectric layers, each characterized by its thickness and complex dielectric function, representing the various physical layers of the samples. 
The optical response of the system was calculated assuming Fresnel boundary conditions at the interface between the layers. 
The thickness of the films was independently measured using profilometry in order to provide a first-order estimation for SE modelling. In addition, the transmission spectrum of 2 at.\% AZO films, deposited on a two-side polished MgO substrate was measured in the 1.00-5.05 eV photon energy range to provide a straightforward evidence for the LSPR response of Au NPs. \\ For the SE modelling, we used WVASE software (J.A. Woollam, Co.), allowing a thorough characterization of the optical constants, film thickness and roughness of the materials involved. Bottom to top, the model included: i) a semi-infinte polished MgO(001) substrate, ii) the AZO film, iii) a roughness layer and iv) an effective layer representative of the Au-NPs deposited on the AZO surface. 
For modelling the optical properties of AZO we resorted to a superposition of Lorentz, Lorentz-gaussian and so-called PSEMI oscillators \cite{herzingerguide1996,patent:5796983}, along with a Drude-type contribution for representing the doping-induced free carriers. 
PSEMI oscillators are parameterized functions widely employed for modelling the optical response of crystalline semiconductors. 
The oscillator parameters of the the undoped ZnO, and the AZO layer, as well as the thickness of all the optical layers were carefully fitted in order to achieve the best agreement between the experimental data and the simulated SE spectra. 
PSEMI oscillators were also employed for the effective modelling of the optical properties of the Au-NP layer on top of the TCO films.
All the layers in the model were isotropic. The existence of any in-plane anisotropy was ruled out by azimuthal-angle-dependent measurements, while the good fitting of the isotropic model applied to measurements collected at different angles of incidence suggests that out-of-plane anisotropy is weak or negligible.

\section{\label{sec:level1}Results and Discussion}

\subsection{\label{sec:level3}Optical properties of AZO films}

\begin{figure}
		\includegraphics[width=8cm]{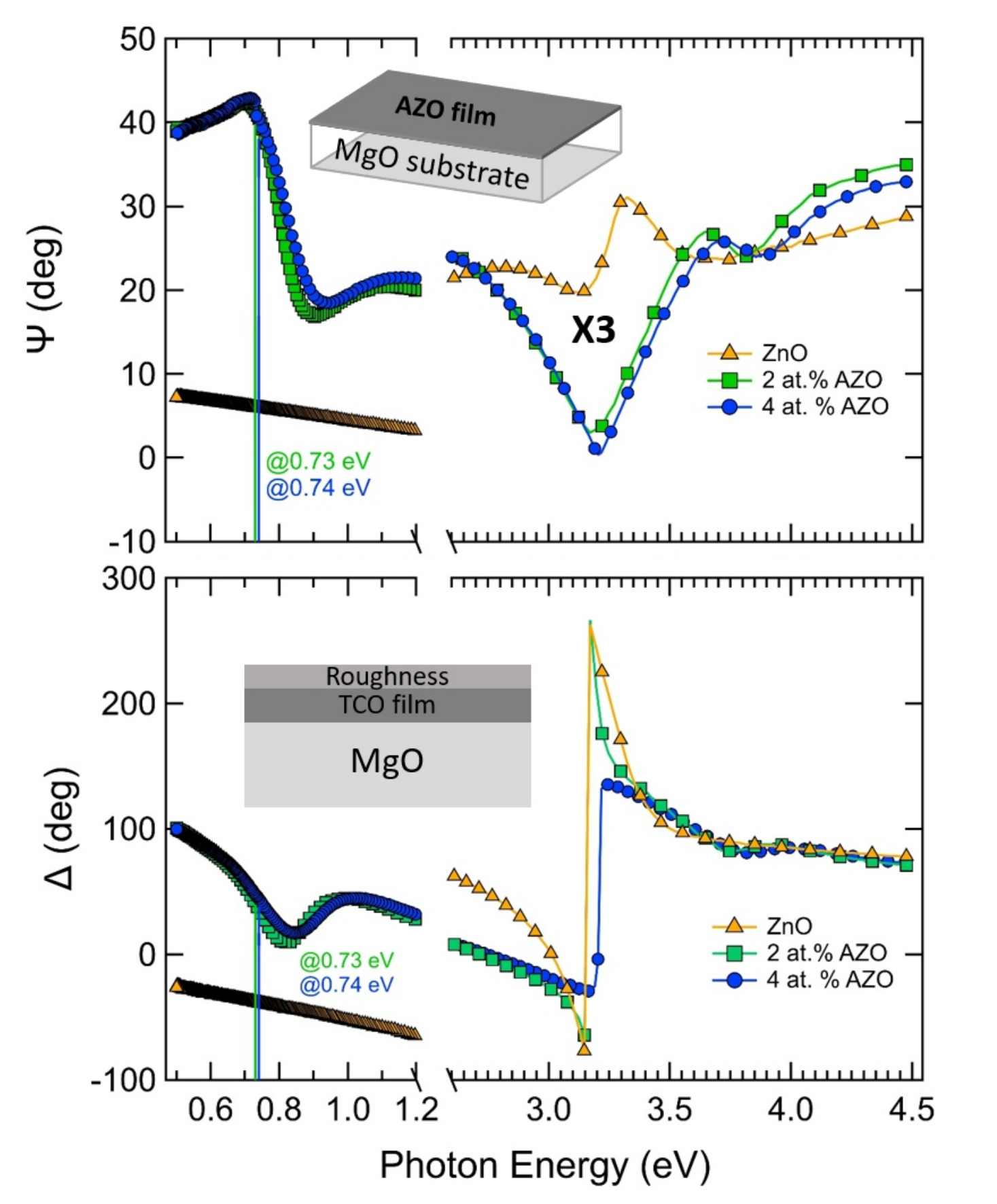}
		\caption{\label{AZO_doping_comparison} $\Psi$ (top) and $\Delta$ (bottom) spectra of ZnO (triangles), 2 at.\% AZO (squares) and 4 at.\% AZO (circles) films, grown on MgO substrates, acquired with incident angle of 65$^{\circ}$. The $\Psi$ spectra above 2.6 eV have been multiplied by a factor of 3 for the sake of clarity. The solid lines in both spectra indicate the plasma frequency for 2 at. \% AZO (green) and 4 at. \% AZO (blue line). The inset on the top image is a representative scheme of the samples under study (AZO film/MgO substrate, 1-side polished). The inset on the bottom image is a representative scheme of the model used for the fit of the experimental ellipsometry data.}
\end{figure}

\begin{figure}[h!]
		\includegraphics[width=9cm]{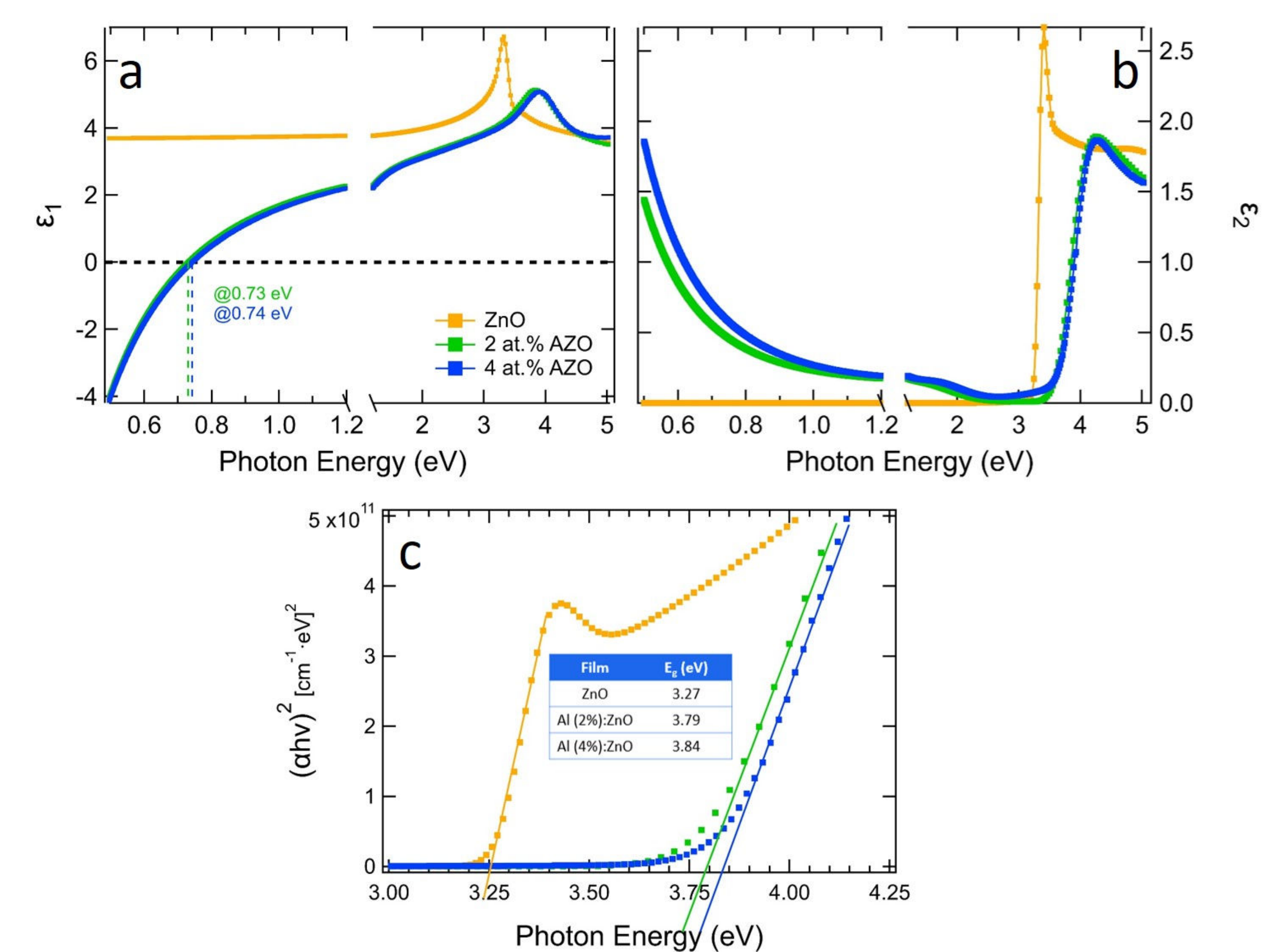}
		\caption{\label{constantsAZO_Taucplot} Real (a) and imaginary (b) part of the dielectric function of AZO films of different doping levels (2, 4 at.\%), as extracted by ellipsometry measurements. Measurements on a ZnO film (black lines) are reported for reference. The dashed lines in $\varepsilon_{1}$ indicate the plasma frequency for 2 at. \% AZO (green) and 4 at. \% AZO (blue line). (c) Squared optical absorption coefficient versus photon energy for AZO films of different doping levels (2,4 at.\%), as well as a ZnO film. The inset table shows the band gap values determined by a linear extrapolation along the absorption edge to the background.}
	\end{figure}

In Figure \ref{AZO_doping_comparison} the ellipsometry spectra $\Psi$ and $\Delta$ of bare and Al-doped (2 at.\% and 4 at.\%) ZnO films, acquired with incident angle of 65$^{\circ}$, are presented in the same graph for comparison. Fits with a mean squared error (MSE) of 2.39, 4.65 and 4.74 were obtained for bare, 2 at.\% doped and 4 at.\% doped film and Al-doped ZnO films, respectively.
Additional spectra  at different incident angles  are shown in Figure S3, along with the best fits obtained in correspondence of the dielectric functions reported in Figure \ref{constantsAZO_Taucplot} and of the morphological parameters (film thickness, roughness) reported below. 
The $\Psi$ and $\Delta$ spectra of undoped ZnO show prominent features close to the bang-gap region, in the near UV, 
and are relatively featureless in the visible-IR regions.
Doped films exhibit a clear evolution of the spectral features in the near-UV, and the appearance of a characteristic features in the near-IR around 0.7 eV, which are attributed to the bulk-plasmon resonance of the AZO films.
The origin and evolution of all these features can be understood by looking at the complex dielectric function $\varepsilon$of the AZO films extracted from these data and presented in Figure \ref{constantsAZO_Taucplot} (the corresponding complex refractive index (\textit{n,k}) are reported in Figure S4). The dielectric function of the AZO films is the physical quantity that carries all the information about the materials’ response (band gap, polarizability, bulk plasmon resonance, Drude contribution) eliminating the thickness dependence.\\
In Figures \ref{constantsAZO_Taucplot}(a) and \ref{constantsAZO_Taucplot}(b) we report the real ($\varepsilon_{1}$) and the imaginary ($\varepsilon_{2}$) part of the dielectric functions of the AZO films as extracted from ellipsometry data. 
A few things stand out from these data.
Starting from the UV range, the optical bandgap was 3.27 eV for pure ZnO films, increasing as a function of doping due to the Burstein-Moss effect, upon which the absorption edge is pushed to higher energies because of the occupation of the bottom of the conduction band \cite{Moss_1954}. 
In Figure \ref{constantsAZO_Taucplot}(c) we report the calculation of the band gap values by means of a Tauc plot. 
Upon 2\% Al-doping the band gap of ZnO increases up to 3.79 eV, while the further increase of the doping pushes the band gap 
to only slightly higher values (3.84 eV). 
This blueshift agrees well with the corresponding blueshift of the UV features in the SE spectra as a function of doping in Figure \ref{AZO_doping_comparison}(top).
The marked excitonic peak at the band-gap edge of ZnO is gradually damped as doping increases.\\
In the visible region (approximately from 2 to 3 eV)  the optical absorption approaches zero as expected for TCO systems. 
The spectral fingerprint of some defect state in the band gap is observable as a deviation from perfect transparency in the visible range. \\
In the near-IR the free-carrier contribution is apparent, with the appearance of a screened plasma frequency at 0.73 eV (0.74 eV) for 2 at. \% AZO (4 at. \% AZO). 
At the plasma frequency, $\varepsilon_2$ values below 0.5 are observed, indicative of high-quality ENZ behaviour.
The crossing of the plasma frequency is actually responsible for the sharp features observed in the SE spectra of Figure \ref{AZO_doping_comparison},
where the dip in the $\Delta$ and the sharp rise in $\Psi$ are the fingerprints of the cross-over from dieletric to metallic behaviour.\cite{ferrera2021}
We point out that the appearance of a plasma frequency and its doping dependence are already qualitatively observable from the raw SE spectra.
The thickness of the films corresponding to the best fit was $110\pm10$ nm, $145\pm15$ nm and $145\pm15$ nm for Al-doped ZnO films (2 at.\% and 4 at.\%, respectively), while the effective optical roughness was found to be approximately $4.0\pm0.1$ nm for ZnO, $6.0\pm0.1$ nm for the 2 at.\% and and $4.6\pm0.2$ nm for the 4 at.\% AZO film. We notice that, whereas the mathematical uncertainty in thickness from the fit was very small ($\pm0.1$ nm), there are other sources of uncertainty (e.g. slight thickness inhomogeneities) that concur in defining a "physical" uncertainty that is larger than the "mathematical" fit uncertainty of the fit.
The slightly different roughness values deduced by SE and AFM are typical of the different lateral scale over which the two techniques assess the sample, and of the "effective" nature of roughness modelling in SE.

From the so-called Drude dielectric function (Eq. \ref{edrude}) and the definition of plasma frequency (Eq. \ref{plasmafrequency}),it
is possible to calculate the density of the free carriers, $N_{e}$. 

\begin{equation}
\varepsilon_{Drude}={\varepsilon_{1}}+i{\varepsilon_{2}}={\varepsilon_{\infty}}- \frac{\omega_{p}^2}{\omega^{2}+i \gamma \omega}
\label{edrude}
\end{equation}

\begin{equation}
\omega_{p}=\sqrt{\frac{N_{e}q^{2}}{m^{*}\varepsilon_{0}}}
\label{plasmafrequency}
\end{equation}

In Eq. \ref{edrude}, $\varepsilon_{\infty}$ is the background permittivity, $\omega$ is the photon frequency, while $\omega_{p}$ is the plasma frequency and $\gamma$ the damping parameter. 
In Eq. \ref{plasmafrequency}, $\varepsilon_{0}$ is the vacuum permittivity, $q$ is the elementary charge and $m^*$ is the effective mass of the charge carriers (for AZO $m^*$=0.27$m_{e}$\cite{effmass}). 
The calculated values of the carrier density of the corresponding films  (2 at.\% and 4 at.\% AZO) were $2.98\cdot10^{20}$  $cm^{-3}$ and $3.91\cdot10^{20}$ $cm^{-3}$ (see Figure S5), respectively, which point out the higher conductivity of the 4 at.\% AZO film, as expected \cite{Benedetti2015,Benedetti2017} and they are in good agreement with literature values.\cite{ZNO2006, ZNO2009}

\subsection{\label{sec:level3}Optical properties of Au nanoparticles on AZO films}

\begin{figure}
		\includegraphics[width=9cm]{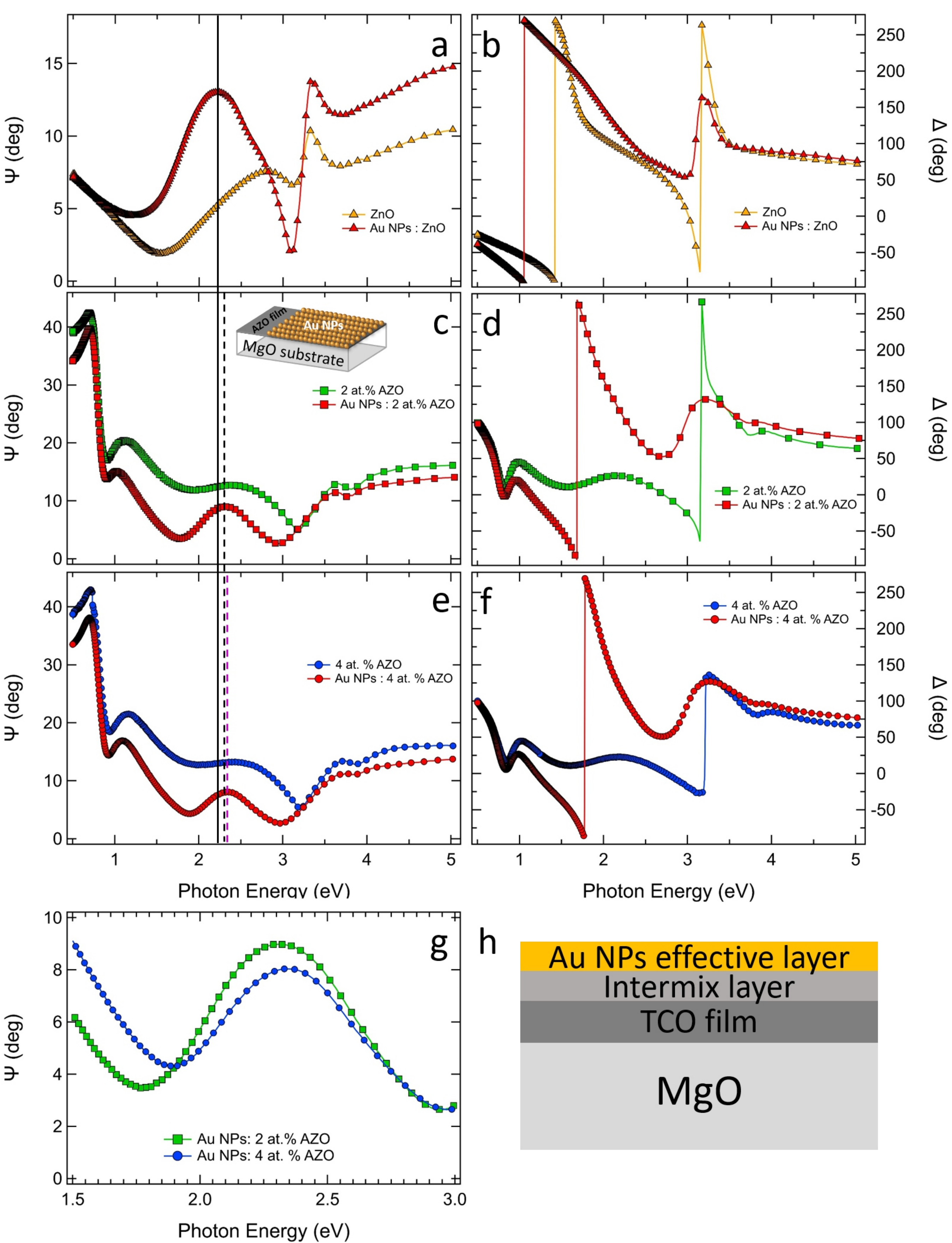}
		\caption{\label{data_w_wo_Au} $\Psi$ (left) and $\Delta$ (right) spectra of ZnO (a, b), Au NPs/2 at.\% AZO (c, d) and Au NPs/4 at.\% AZO (e, f) films, with and without Au NPs on top, acquired with incident angle of 65$^{\circ}$. In all spectra, the red markers represent the data of the Au NPs-TCO system. Three lines were placed on the $\Psi$ spectra in order to point out the peak, related to LSPR of Au NPs. The solid line was set at the peak of ZnO spectrum while the dashed lines indicate the peaks of Au NPs/2 at.\% AZO (black dashed line) and Au NPs/4 at.\% AZO (purple dashed line) spectra. (g) $\Psi$ spectra of Au NPs/2 at.\% AZO (squares) and Au NPs/4 at.\% AZO (circles) films, acquired with incident angle of 65$^{\circ}$. Inset in Figure \ref{data_w_wo_Au}(c): A representative scheme of the samples under study (Au NPs/AZO film/MgO, substrate 1-side polished. (h) A representative scheme of the model used for the fit of the experimental ellipsometry data.}
\end{figure}

The SE spectra of the Au-NP/AZO system are reported in Figure \ref{data_w_wo_Au}. 
In the same graph, we report the bare-substrate spectra for the sake of comparison. 
The SE spectra of the Au NPs/TCOs in different angles of incidence, along with the fit curves are shown in Figure S6. 
The SE spectra of the hybrid system are different with respect to their bare counterpart, due to the introduction of an additional dispersive layer on top of the system.
A common feature of the SE spectra of Figure \ref{data_w_wo_Au} seems to be the appearance of a peak around 2.1 eV in $\Psi$, intuitively related to the LSPR of Au NPs on top of the TCO surface 
(see also the transmission spectra of bare and NP-decorated AZO films reported in Figure S7, where a clear fingerprint of the LSPR is seen at that energy). 
By comparing the ellipsometry spectra of Au NPs on bare and Al-doped (2\% and 4\%) ZnO (Figure \ref{data_w_wo_Au}) a slight blue shift of such a structure is observed as a function of increasing doping. We observe a slight decrease of the peak width as the doping of the TCO films is increased from 2\% to 4\% Figure \ref{data_w_wo_Au}(g). In order to shed more light on these observations, a dielectric modelling of the Au-NP/AZO system is required.

\begin{figure}
		\includegraphics[width=9cm]{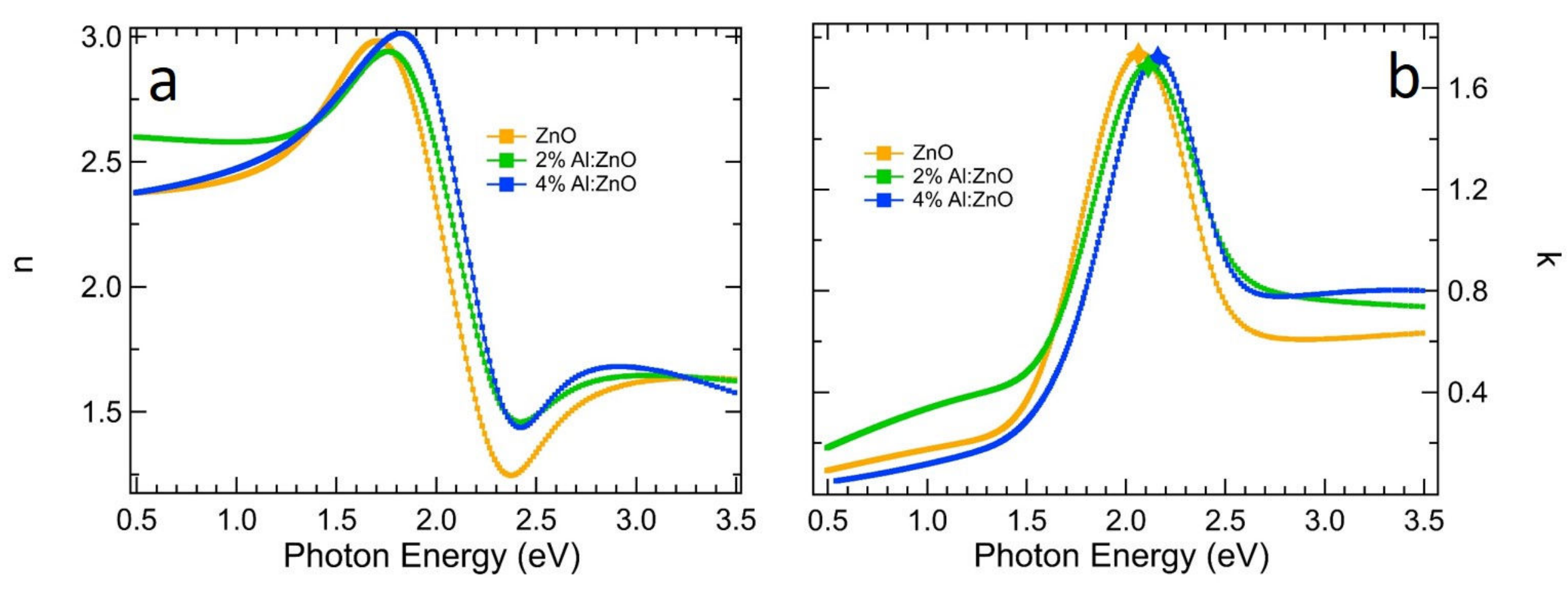}
		\caption{\label{n_k_Au_AZO} Refractive index, n (a) and extinction coefficient, k (b) of the effective Au-NP layer on bare ZnO and AZO films (2\% and 4\%), as extracted by spectroscopic ellipsometry. Markers on the $k$ peak were placed for the sake of clarity of the LSPR blueshift.}
\end{figure}

To fit the new spectra and extract the optical response of the Au-NP layer, the SE model was accordingly modified to include a new dielectric layer, representative of the effective dielectric function of an Au-NP assembly\cite{Choy1999,Decker2018}, which was approximated as dielectrically isotropic.\cite{Anghinolfi2011} Fits with MSE of 10.25, 6.82 and 25.55 for hybrid systems of Au NPs on top of bare and Al-doped (2 at.\% and 4 at.\%) ZnO films were achieved, respectively.
Figure \ref{n_k_Au_AZO} reports the refractive index \textit{n} (a) and extinction coefficient \textit{k} (b) of the effective Au-NP layer on bare ZnO and AZO films (2 and 4 at.\%), grown on MgO substrates, as extracted from the real ($\varepsilon_{1}$) and imaginary ($\varepsilon_{2}$) part of the dielectric function, corresponding to the best fit (Figure S8).

The extinction coefficient, \textit{k}, of the Au-NP layer, extracted via the best fitting of the SE spectra, shows a resonance at 2.06 eV for Au on ZnO, 2.11 eV for Au on 2 at.\% AZO and 2.16 eV for Au on 4 at.\% AZO. 
This resonance, the spectral fingerprint of the LSPR of Au NPs, exhibits a clear blueshift
as a function of increased doping of the TCO film, as actually observed in the raw SE data.
We point out that the effective dielectric function of the Au-NP layer extracted from SE also allows to reproduce, with very minor changes, the transmission data (see e.g. the case of Au NPs/2 at.\% AZO reported in Figure S9), thereby supporting the analysis results.

At this point, it is worth mentioning that there is rather conclusive evidence that the NPs formed on top of the TCOs are separated into islands. If the Au atoms still formed a continuous film, we would not witness the presence of a LSPR in the optical data (both transmission and ellipsometry). The optical properties would be completely different with respect to our observations whereas for the electronic properties, a conductive layer would be formed on top of the AZO, contributing to the overall conductivity of the system.

In Figure \ref{n_k_Au_AZO}(b) a decrease of Au plasmon linewidth is confirmed going from 2\% to 4\% samples. We ascribe this phenomenon to the lower dissipation in the substrate in correspondence of the LSPR of 4\%-doped AZO.


The evolution of the LSPR on the differently-doped substrates can originate from a variety of effects.
Size and shape of the particles are one such effect, but since AFM images provide no clearcut evidence of a systematic variation
of these parameters, we are inclined to rule them out. As a matter of fact, the mean radius of Au NPs on top of bare and Al-doped (2\% and 4\%) ZnO films as extracted from AFM images analysis was found to be $10.9\pm0.2$ nm, $11.1\pm0.2$ nm and $10.6\pm0.2$ nm, respectively. Due to this analysis, no size trend is observed, so the LSPR shift cannot directly stem from these variations in size. Moreover, the mean radius of the NPs lies in a size range where the size dependence of the LSPR is extremely weak\cite{SizeandTemperatureDependence}  meaning that the shift in LSPR cannot be attributed to the small variations in the NPs size.
Next, we consider the effect of particle environment, {\em i.e.} the interactions with the substrate, dividing them in two categories, namely the purely "dielectric" interactions 
(shift of the LSPR due to the different polarizability of the surrounding material), and electronic interactions, {\em i.e.}
that imply a transfer of charge between the materials.

In a simple approximation, for a given metal NP with fixed size and shape, there is a linear relation connecting the dielectric constant of the environment at the resonance frequency and the LSPR frequency.\cite{environment}
In our case, this is only qualitatively verified:
indeed, the blueshift of the LSPR when going from ZnO to 2\%-doped AZO can be rationalized based on the 
corresponding decrease of $\varepsilon_1$ of the substrate in the spectral region of the LSPR.
However, when going from the 2\%-doped AZO to the 4\%-doped AZO we observe a further blueshift not motivated by a corresponding
variation of the dielectric function. 

An appealing hypothesis for this is related to the occurrence of a charge transfer between the two materials. Hot electrons play a role in several hybrid systems that include plasmonic materials. On these basis, we cannot rule out their influence on our observations, even though we do not have the possibility to single them out. In general, however, we understand our observations in terms of a net transfer of electrons towards the Au NPs.
Indeed, the increase of chemical potential in the AZO due to the doping may promote a charge accumulation 
in the Au NPs, which would lead to a blueshift of the LSPR even in absence of concomitant effects (shape, size, dielectric environment). Taking into account the work function of bare and Al-doped (2 at. \% and 4 at. \%) ZnO as extracted from Ultraviolet Photoelectron Spectroscopy measurements (3.96 eV, 4.17 eV and 4.51 eV, respectively) knowing that the work function of gold is 5.3 eV and since ZnO is an n-type semiconductor we could assume that the contact between Au NPs and ZnO is a Schottky contact.\cite{Catellani2014} However, the difference between the work functions of gold and doped ZnO films are relatively small. Furthermore the presence of occupied states at Fermi level and the ability in screening charges in the conductive AZO has shown several times a Ohmic contact.\cite{Ahmadi2021} Therefore, distinguishing the type of metal/semiconductor junction can not be accomplished just by knowing the work function of the materials involved, because the role of the interface as well as the presence of defects can affect the type of contact.\cite{Brillson2011,ZnO2005}. Assuming that charge injection from the AZO substrate to the Au NPs is responsible for the blueshift of the plasmon resonance, according to our calculations the relative variations of the carrier density of gold corresponding to a plasmon resonance shift from 2.06 eV to 2.11 eV and 2.16 eV should be $3.3\cdot10^{21}$  $cm^{-3}$ and $6.7\cdot10^{21}$ $cm^{-3}$, respectively. 
Considering that the Au coverage corresponds to a nominal 3 nm thickness, this leads to an average transferred charge to the NPs of about 0.05 e/atom and 0.11 e/atom for 2 at.\% and 4 at.\% AZO films, respectively. 
These values are consistent, in absolute value, to what happens in other metal NPs/ZnO\cite{Benedetti2020,Chernysheva2018} and similar Au NPs/oxide systems\cite{mitsuhara2010,Torelli2009} and can be sustained by the substrate due to the small volume of Au NPs with respect to the film volume. 
Increasing further the dopant concentration above the optimal doping condition (4 at. \%) introduces Al in interstitial sites, depopulation of the conduction band and new defect states in the band gap \cite{Benedetti2015,Benedetti2017}. The larger degree of complexity would in turn make it more difficult to understand and model charge transfer phenomena at the interface. This clearly represents an upper limit, assuming the doping-dependent variation of the polarizability of the substrate plays no role in the blueshift. It is possible that both effects play a concomitant role in determining the actual value of LSPR shift. 
In this respect, we speculate that in analogous systems with a higher ratio between Au NP volume and TCO volume, some of these effects could reflect on a variation of the optical response of the substrate following the deposition of plasmonic particles, promoting an intertwinned TCO/NPs relationship, that can potentially be exploited to tailor the ENZ regime as well.

\newpage
\section{\label{sec:level1}Conclusions}
In conclusion, we reported a spectroscopic ellipsometry investigation of the dependence of the LSPR of metallic NPs on the doping of a TCO film (here, Al-doped ZnO films).
We investigated first the evolution of the dielectric properties of the substrate as a function of Al doping, recording an increase of the TCO band gap as a function of increasing doping  due to the Moss-Burstein effect, and the apperance of Drude tail distribution due to the presence of free carriers. 
In addition, we showed that the LSPR of Au NPs deposited on the TCO blueshifted as a function of increasing doping.
Such a blueshift cannot be simply understood in terms of a variation of the dielectric environment of the plasmonic nanoparticles,
hence we suggested that a doping-dependent charge transfer between the substrate and the NPs is responsible for the effect.
This could have interesting implications in terms of either passively or actively tuning the optical response of hybrid plasmonic/TCO systems,
exploitable for a broad range of energy/environmental applications, such as in light harvesting and photocatalytic devices.

\newpage
\section*{Conflicts of interest}
There are no conflicts to declare.

\subsection*{Supporting Information}

AFM images of bare ZnO and AZO films of different doping levels, AFM images of Au NPs deposited on top of ZnO and AZO films, SEM and AFM images of reference ZnO films, Ellipsometric spectra $\Psi$ and $\Delta$ and optical properties of bare and Al-doped (2 at.\% and 4 at.\%) ZnO films, Ellipsometric spectra of the Au-NP layer deposited on top of ZnO and AZO films, Transmission spectra of AZO/MgO and Au NPs/AZO films, Modelling of the transmission measurements of Au NPs/AZO films and the extracted optical properties.

\medskip
\section*{Acknowledgements}

This project has received funding from the European Union's Horizon 2020 research and innovation programme under the Marie Sk\l odowska-Curie grant agreement N$^{\circ}$799126. 

\section*{Data Availability}
The data that supports the findings of this study are available within the article [and its supplementary material].


\providecommand{\noopsort}[1]{}\providecommand{\singleletter}[1]{#1}%
\providecommand{\latin}[1]{#1}
\makeatletter
\providecommand{\doi}
  {\begingroup\let\do\@makeother\dospecials
  \catcode`\{=1 \catcode`\}=2 \doi@aux}
\providecommand{\doi@aux}[1]{\endgroup\texttt{#1}}
\makeatother
\providecommand*\mcitethebibliography{\thebibliography}
\csname @ifundefined\endcsname{endmcitethebibliography}
  {\let\endmcitethebibliography\endthebibliography}{}

\newpage
\begin{figure}
\medskip
\includegraphics[width=\linewidth]{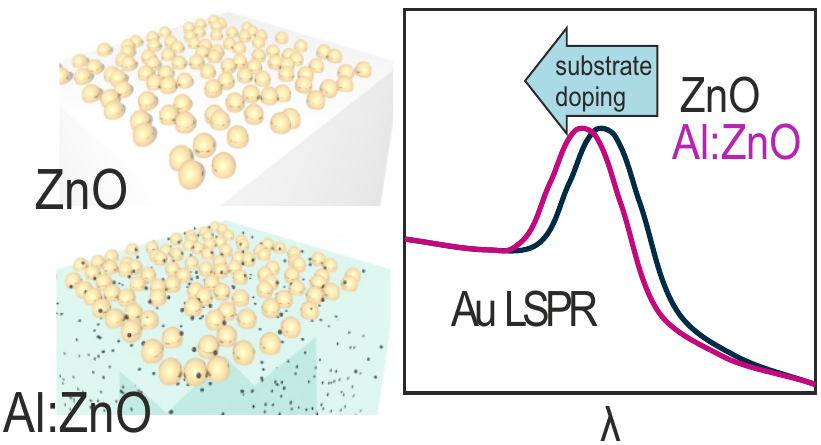}
  \medskip
  \caption*{TOC Graphic}
\end{figure}

\end{document}


{\color{red}{This article may be downloaded for personal use only. Any other use requires prior permission of the author and ACS Publishing.  This article appeared in (M. Sygletou,The Journal of Physical Chemistry C (2022)) and may be found at (https://10.1021/acs.jpcc.1c07567).}}

\maketitle



\begin{figure*}[h!]
		\includegraphics[width=14cm]{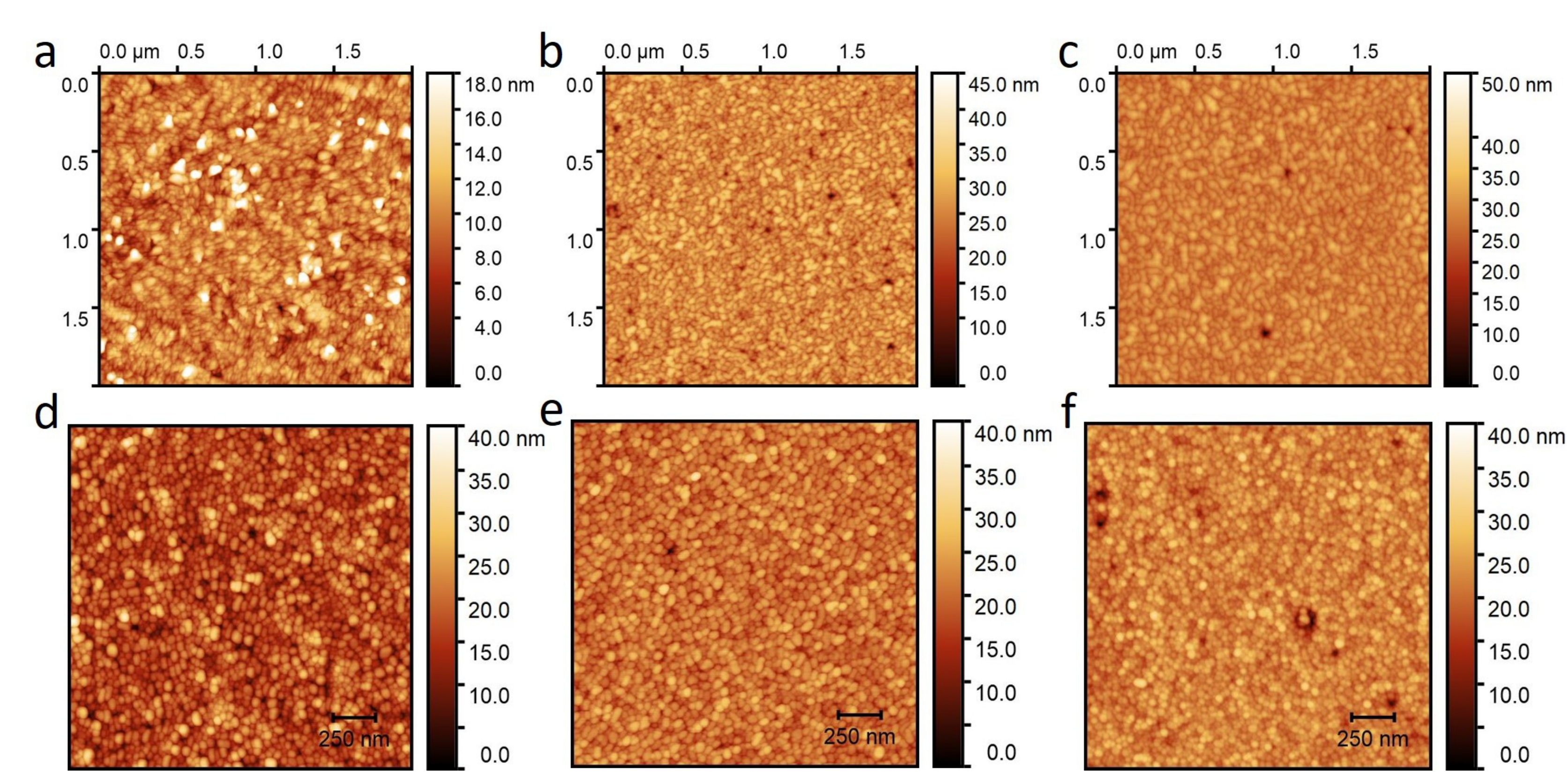}
\renewcommand*{\thefigure}{S\arabic{figure}}
		\caption{\label{AFM_all}Top: AFM images of ZnO (a), 2 at.\% AZO (b) and 4 at.\% AZO (c) films of 150 nm thickness, grown on MgO substrates. Bottom: AFM images of Au NPs deposited on top of ZnO (d), 2 at.\% AZO (e) and 4 at.\% AZO (f) films, grown on MgO substrates.}
	\end{figure*}

\begin{figure*}[h!]
		\centering
		\includegraphics[width=18cm]{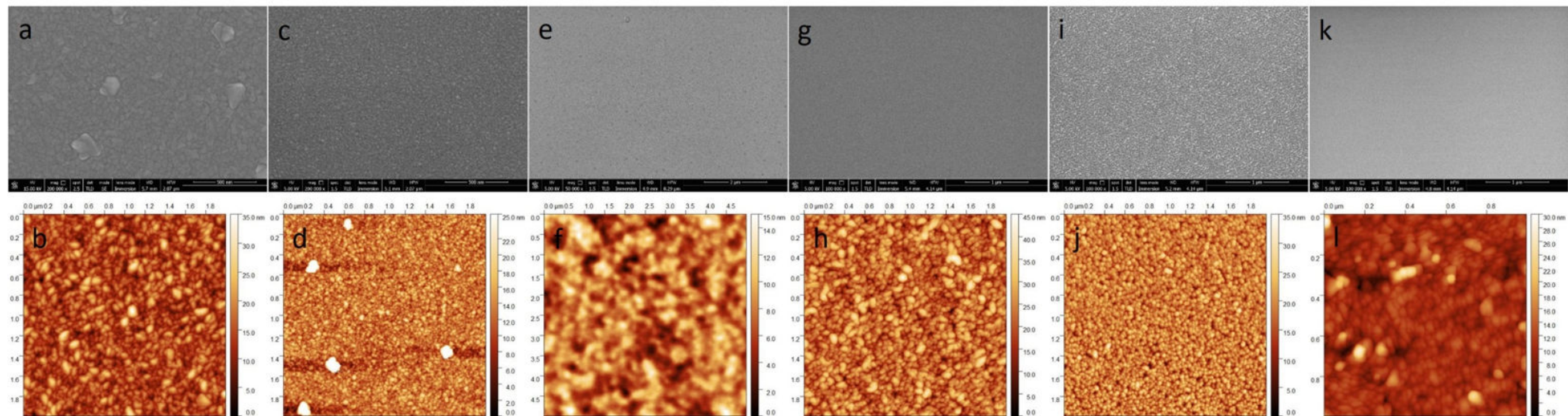}
\renewcommand*{\thefigure}{S\arabic{figure}}
		\caption{\label{ZnO_AFM}Typical SEM (top) and AFM (bottom) images of ZnO films on Si/SiO${_2}$ substrate, fabricated with different deposition parameters. (a,b) ZnO film, fabricated with the deposition parameters, used as reference (Deposition of 300 nm ZnO at room temperature, in the presence of Argon, deposition rate of 1.1Å/s, pressure of 5 mTorr). In the other depositions only one parameter was changed, leaving the remaining parameters unaltered. (c,d) Deposition of ZnO film of thickness 100 nm, (e,f) deposition of ZnO film at $300^{\circ}$C, (g,h) deposition of ZnO film with rate of 0.5A/s and (i,j) deposition of ZnO film under the prescence of O${_2}$. (k,l) Depostion of 100 nm ZnO film at  $300^{\circ}$C.}
	\end{figure*}

\newpage
AFM images of bare ZnO and AZO films of different doping levels (2\% and 4\%), obtained in the same deposition conditions, are shown in Figure \ref{AFM_all} (a-c). AFM images of Au NPs deposited on top of ZnO and AZO films, are also shown in Figure \ref{AFM_all} (d-f). Representative SEM and AFM images of the parametric study for the optimization of the surface morphology and roughness of bare ZnO films on MgO substrates, by modifying the deposition parameters, are presented in Figure \ref{ZnO_AFM}. \par

\begin{figure*}[h!]
		\includegraphics[width=12cm]{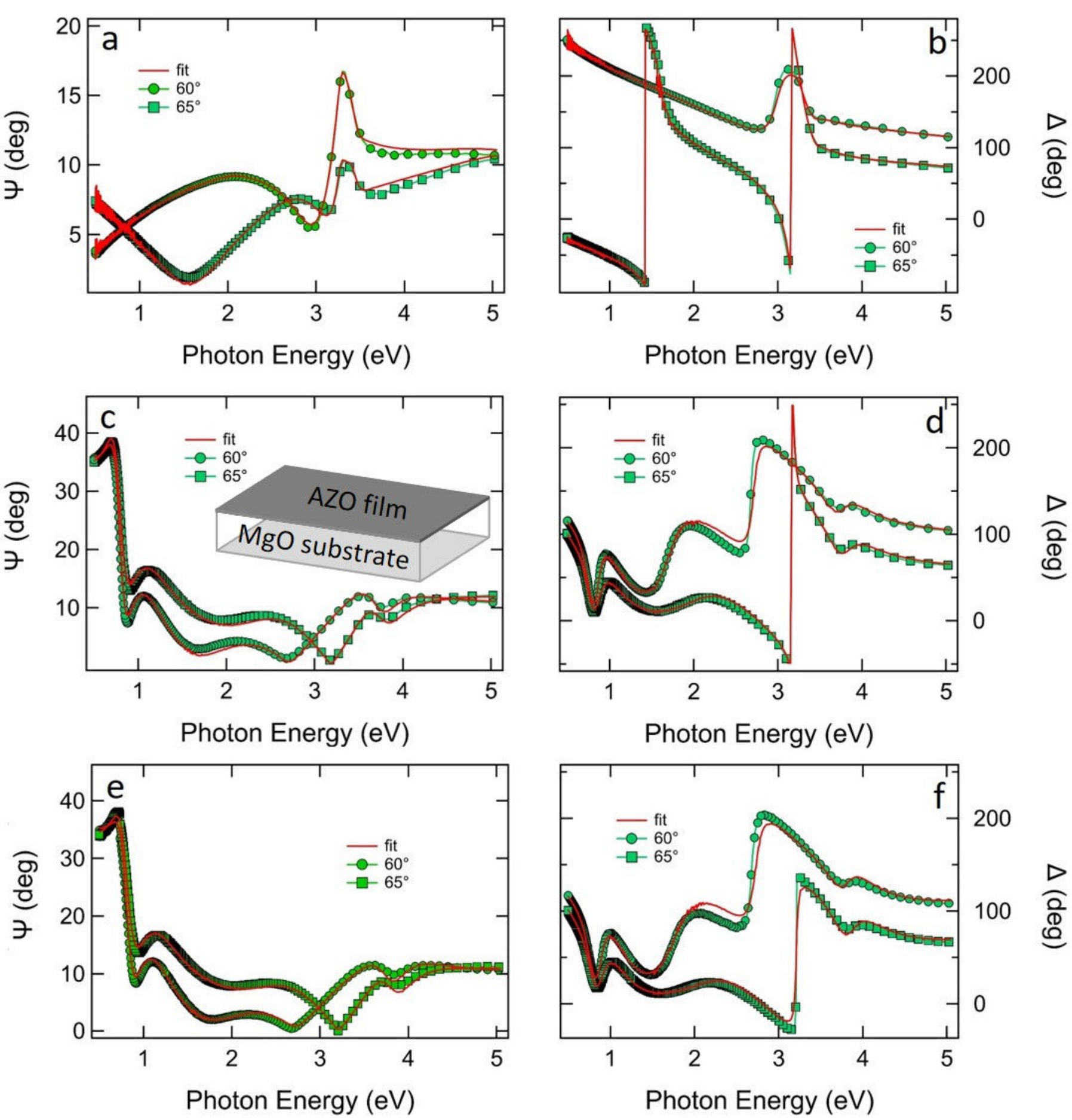}
\renewcommand*{\thefigure}{S\arabic{figure}}
		\caption{\label{AZOpsi_delta_Newsetup} $\Psi$ (left) and $\Delta$ (right) spectra of ZnO (a, b), 2 at.\% AZO (c, d) and 4 at.\% AZO (e, f) films, grown on MgO substrates, acquired with incident angle of 60$^{\circ}$(circles) and 65$^{\circ}$(squares). Lines represent the best fit to the experimental data. The inset of Figure \ref{AZOpsi_delta_Newsetup}(c) is a representative scheme of the samples under study (AZO film/MgO substrate, 1-side polished).}
\end{figure*}

\begin{figure*}[h!]
\centering
		\includegraphics[width=12cm]{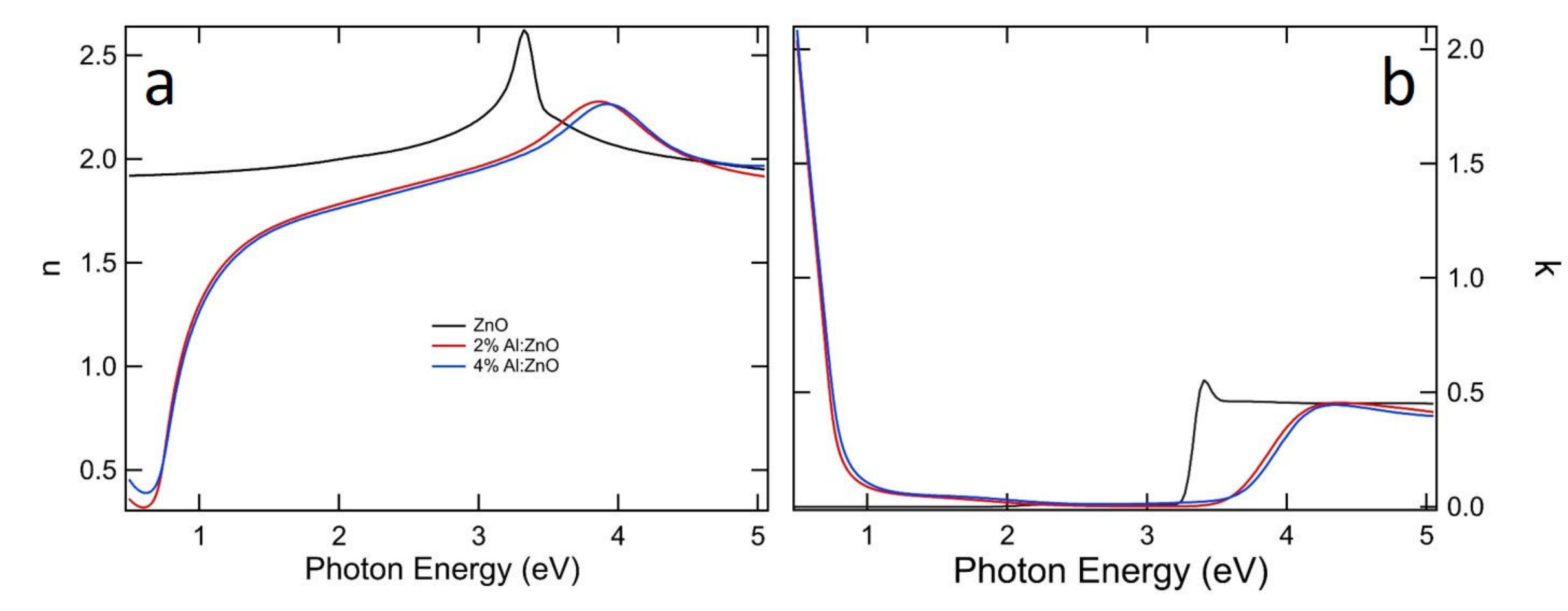}
\renewcommand*{\thefigure}{S\arabic{figure}}
		\caption{\label{constantsAZOnk} Refractive index, n, (a) and extinction coefficient, k, (b) of AZO films of different doping levels (2, 4\% at.). The optical properties of a ZnO film (black lines) are reported for reference.}
	\end{figure*}

In Figure \ref{AZOpsi_delta_Newsetup} the ellipsometric spectra $\Psi$ and $\Delta$ of bare and Al-doped (2 at.\% and 4 at.\%) ZnO films, acquired with incident angles of 60$^{\circ}$ and 65$^{\circ}$, are shown, top to bottom.
Green symbols correspond to experimental points, while red lines represent the best fit obtained in correspondence of the optical properties reported in Figure \ref{constantsAZOnk} and of the morphological parameters (film thickness, roughness) reported in the manuscript.
\par

\begin{figure*}[h!]
		\centering
		\includegraphics[width=8cm]{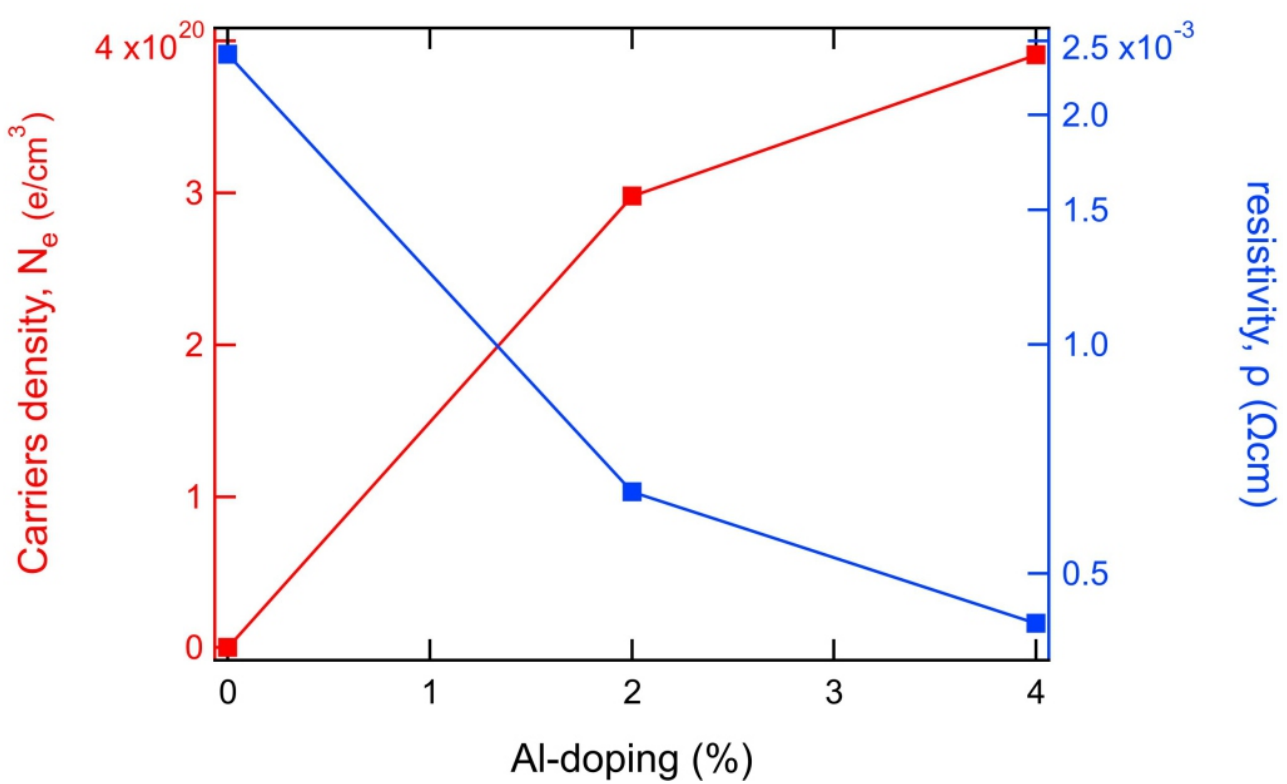}
\renewcommand*{\thefigure}{S\arabic{figure}}
		\caption{\label{N_ro} Carriers density ($N_{e}$) and resistivity ($\rho$) of bare and Al-doped ZnO films.}
	\end{figure*}

In Figure \ref{N_ro} the carriers density ($N_{e}$) and resistivity ($\rho$) of bare and Al-doped (2 at.\% and 4 at.\%) ZnO films, as extracted from SE, are reported.

\begin{figure*}[h!]
		\includegraphics[width=12cm]{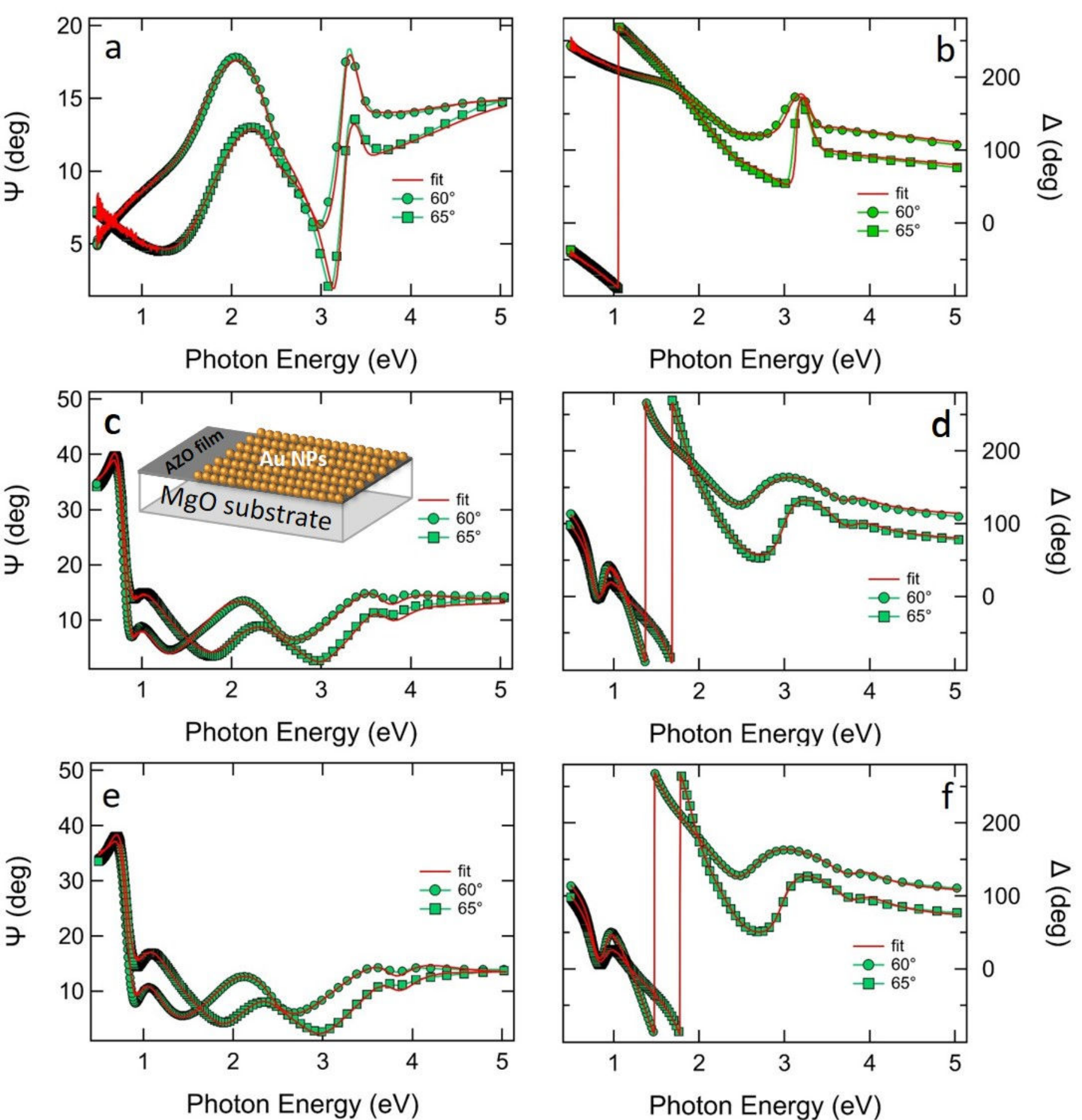}
\renewcommand*{\thefigure}{S\arabic{figure}}
		\caption{\label{AZO_Au_psi_delta_newsetup} $\Psi$ (left) and $\Delta$ (right) spectra of Au NPs/ZnO (a, b), Au NPs/2 at.\% AZO (c, d) and Au NPs/4 at.\% AZO (e, f) films, acquired with incident angle of 60$^{\circ}$(circles) and 65$^{\circ}$(squares). Lines represent the best fit to the experimental data. In Figures \ref{AZO_Au_psi_delta_newsetup}(c) and (d) the ellipsometric parameters $\Psi$ and $\Delta$ of 2 at.\% AZO film, acquired with incident angle of 60 $^{\circ}$ (blue circles) are shown for comparison. The inset of Figure \ref{AZO_Au_psi_delta_newsetup}(c) is a representative scheme of the samples under study (Au NPs/AZO film/MgO, substrate 1-side polished).}
\end{figure*}

\begin{figure*}[h!]
		\centering
		\includegraphics[width=12cm]{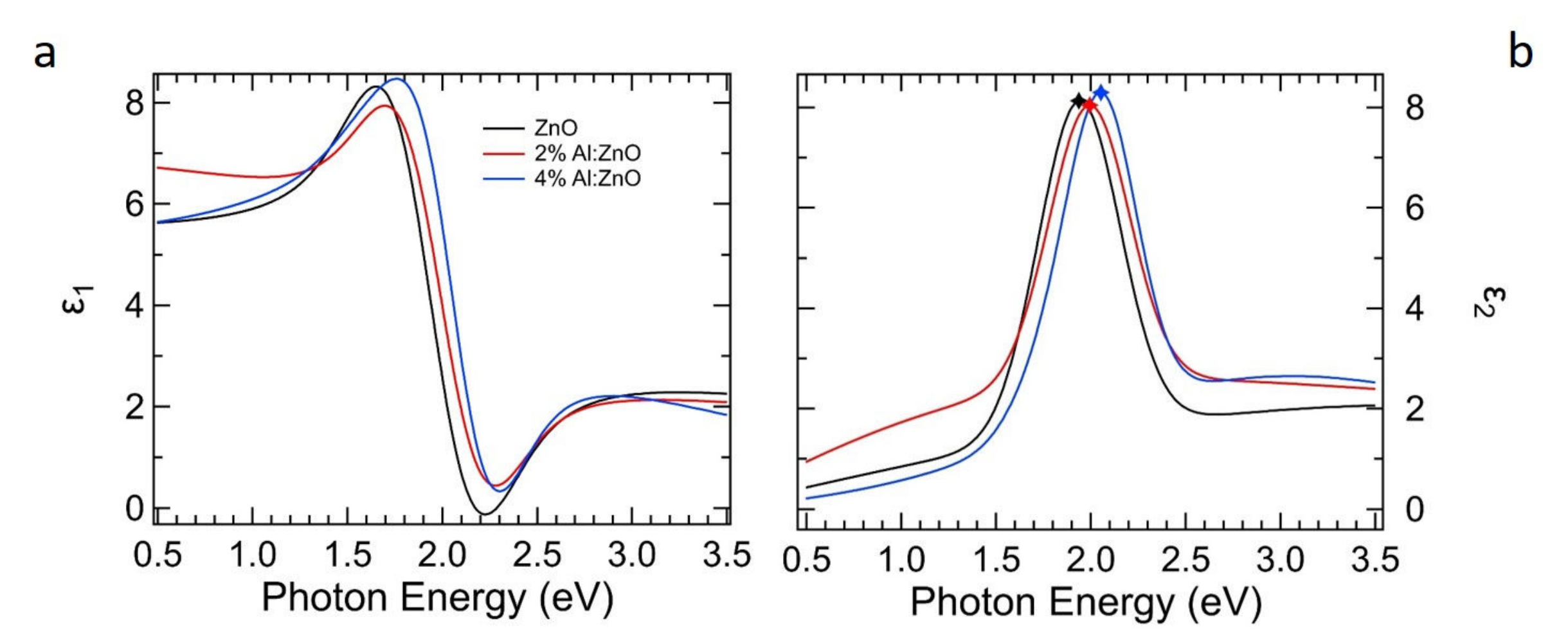}
\renewcommand*{\thefigure}{S\arabic{figure}}
		\caption{\label{Au_AZO_images} Real (a) and imaginary (b) part of the dielectric function of Au NPs on bare ZnO and AZO films (2 and 4 at.\%), as extracted by spectroscopic ellipsometry. Markers on the $\varepsilon_{2}$ peak were placed for the sake of clarity of the LSPR blueshift.} 
	\end{figure*}

The SE spectra $\Psi$ and $\Delta$ of  the Au-NP layer deposited on top of bare and Al-doped (2 at.\% and 4 at.\%) ZnO films, acquired with incident angles of 60$^{\circ}$ and 65$^{\circ}$, are shown, top to bottom, in 
Figure \ref{AZO_Au_psi_delta_newsetup}.
Green symbols correspond to experimental points, while red lines represent the best fit obtained in correspondence of the dielectric function reported in Figure \ref{Au_AZO_images}.\par


The intermix layer between ZnO (bare and Al-doped) and Au NPs is an Effective Medium Approximation (EMA) layer that models interfacial mixing or “interface” roughness by mixing the layers above (Au NPs) and below (AZO) the interface in 50:50 Bruggeman EMA. The thickness of the intermix layers between the bare and Al-doped (2 at. and 4 at. \%) ZnO films and Au NPs was and $7.0 \pm 0.3$ nm,$7.1 \pm 0.1$ nm and $7.6 \pm 0.6$ nm, respectively.

\begin{figure*}[h!]
		\centering
		\includegraphics[width=8cm]{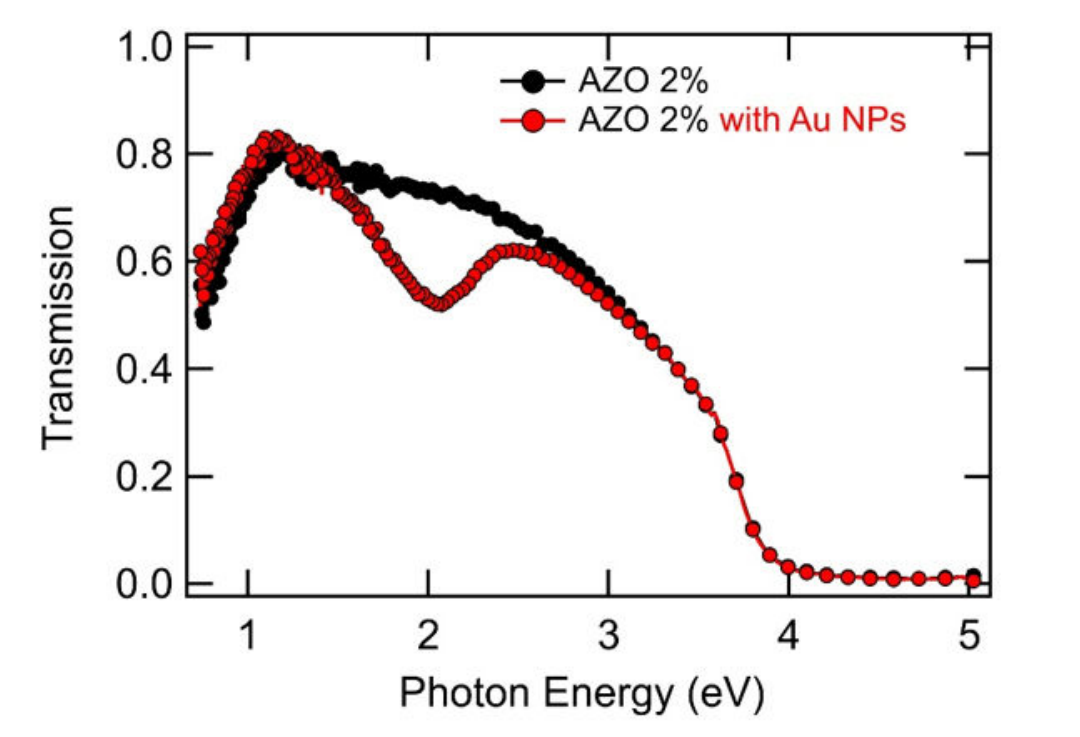}
\renewcommand*{\thefigure}{S\arabic{figure}}
		\caption{\label{Transmission} Transmission spectra of 2 at.\% AZO films with (red markers) and without (black markers) gold NPs on top.}
	\end{figure*}

In Figure \ref{Transmission} we report the transmission spectra of AZO/MgO films, bare (black markers) and following the deposition of Au NPs (red markers). There we observe clearly the effect of the LSPR, which is less obvious in SE. The transparent nature in the visible region of 2 at.\% AZO films fabricated in this work, is shown while the transmission dip around 600 nm in the red spectrum is the fingerprint of the LSPR.

\begin{figure*}[h!]
		\centering
		\includegraphics[width=17cm]{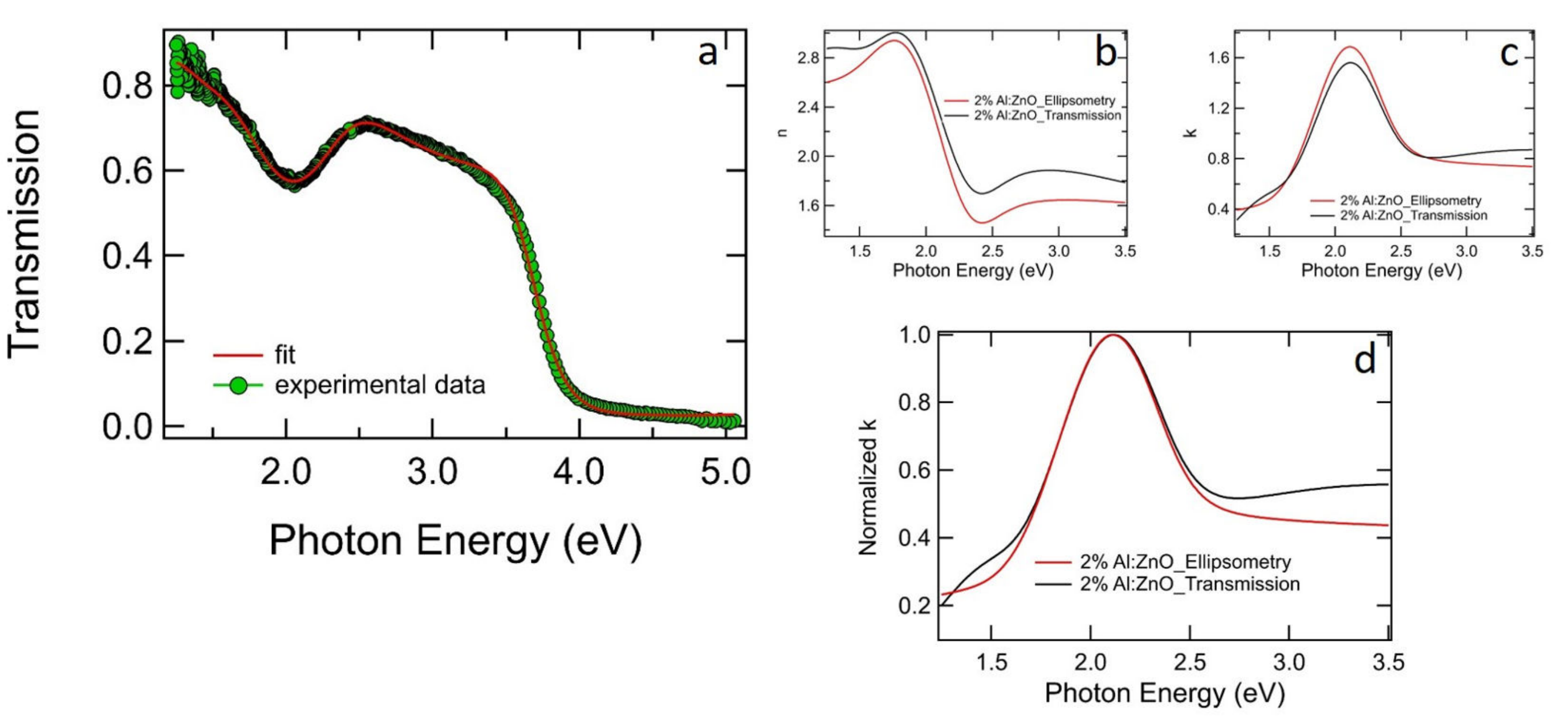}
\renewcommand*{\thefigure}{S\arabic{figure}}
		\caption{\label{T_E} (a)Transmission spectrum of  Au NPs on 2 at.\% AZO film of 100 nm thickness, grown on 2-side polished MgO substrate. Circles represent the experimental data while lines the theoretical curve, calculated from the dielectric parameters obtained from ellipsometry measurements on 1-side polished MgO substrate as in Figure 5 of the manuscript. Effective refractive index, n, (b) and extinction coefficient, k, (c) of the Au-NP layer deposited on the 2 at.\% AZO film on 2-side polished MgO substrate (black lines). The optical properties of Au NPs/2 at.\% AZO on 1-side polished MgO (red lines) are shown for comparison. (d) Normalized extinction coefficient of the two hybrid systems.}
	\end{figure*}

The optical constants of Au NPs/AZO films extracted from ellipsometry were applied for the modelling of transmission measurements. The system was modelled replacing the MgO 1-side polished with a 2-side polished MgO substrate. In Figure \ref{T_E}(a) the experimental data from the transmission measurements are shown (green markers) along with the fit data (red lines) which are the outcome of the modelling. The respective refractive index and extinction coefficient that came out of the modelling of this system are also shown in Figure \ref{T_E}.\par

\newpage